\documentclass[aps,prl,showpacs,twocolumn]{revtex4}

\usepackage{amsmath,epsfig,amsfonts,epsfig,graphicx,
}

\begin{document}

\title{Highly Nonlinear Solitary Waves in Periodic Dimer Granular Chains}
\author{Mason A. Porter$^{1}$, Chiara Daraio$^{2*}$,
Eric B. Herbold$^3$, Ivan Szelengowicz$^{4}$, and P. G. Kevrekidis$^5$}
\affiliation{ 
$^1$Oxford Centre for Industrial and Applied Mathematics, Mathematical Institute, University of Oxford, OX1 3LB, United Kingdom \\
$^2$Graduate Aeronautical Laboratories (GALCIT) and Department of Applied Physics,
California Institute of Technology, Pasadena, CA 91125, USA \\
$^3$ Department of Mechanical and Aerospace Engineering, University of California at San Diego, La Jolla, California 92093-0411, USA \\
$^4$ D\'{e}partement de M\'{e}canique, Ecole Polytechnique, 91128 Palaiseau cedex, France \\ 
$^5$Department of Mathematics and Statistics, University of Massachusetts, Amherst MA 01003-4515, USA }

\begin{abstract}
We report the propagation of highly nonlinear solitary waves in heterogeneous, periodic granular media using experiments, numerical simulations, and theoretical analysis.  We examine periodic arrangements of particles in experiments in which stiffer/heavier beads (stainless steel) are alternated with softer/lighter 
ones (polytetrafluoroethylene beads).  
We find excellent agreement between experiments and numerics in a model with Hertzian interactions between adjacent beads, which in turn agrees very well with a theoretical analysis of the model in the long-wavelength regime that we derive for heterogeneous environments and general bead interactions.  Our analysis encompasses previously-studied examples as special cases and also provides key insights
on the influence of the dimer lattice on the properties (width and propagation speed) of the obtained highly nonlinear wave solutions. 
\end{abstract}


\maketitle


Over the past several years, the highly nonlinear dynamic response of 
granular materials has drawn increased attention from the 
scientific community 
\cite{nesterenko1,coste97,nesterenko2,dar05,dar05b,dar06,hascoet00,hinch99,man01,hong01,hong02,hong05,job05,doney06,ver06,sok07}.
The corresponding theory developed for uniform lattice 
systems \cite{nesterenko1} supports the formation of a novel type of 
wave in materials, setting a paradigm for the design and creation of 
systems with unprecedented properties.  A simple setup for the study of highly nonlinear dynamics in solids is provided by one-dimensional (1D) granular media consisting of chains of interacting spherical particles that deform elastically when they collide.  The broad interest in such systems has arisen because they possess qualitatively different features from weakly nonlinear systems.  For example, their solitary-wave solutions have a finite support that is independent of their 
amplitude \cite{nesterenko1}, providing perhaps the most experimentally 
tractable application of the notion of ``compactons'' \cite{rosenau1}.  
There have also been a number of recent studies on the effects of defects (i.e., inhomogeneities, particles with different masses, etc.) in such systems, allowing the observation of interesting physical responses such as fragmentation, anomalous reflections, and energy trapping \cite{hascoet00,ver06,hinch99,man01,hong02,hong05,job05,dar06,doney06}.  Moreover, chains of granular media have been shown to be highly tunable \cite{nesterenko1,nesterenko2,coste97} and have the potential to be used in many engineering applications -- including shock and energy absorbing layers \cite{dar06,hong05,doney06,ver06}, sound focusing devices (tunable acoustic lenses and delay lines), sound absorption layers, and sound scramblers \cite{dar05,dar05b}.

A class of  phenomena for which 1D chains of granular media provide an ideal setting concerns the interplay between nonlinearity and periodicity.  The study of nonlinear oscillator chains has a 
time-honored history, originating with the Fermi-Pasta-Ulam problem \cite{focus,Ford,linden03}.  Its applications arise in numerous areas of physics, including coupled waveguide arrays and photorefractive crystals in 
nonlinear optics \cite{photon,efrem1}, Bose-Einstein condensates in optical lattices in atomic physics \cite{morsch1}, and DNA double-strand dynamics in biophysics \cite{peyrard}.  A particular theme that often arises in this context is that of ``heterogeneous'' versus ``uniform'' lattices.  Here, we focus on the prototypical heterogeneity of ``dimers'' (i.e., chains of beads made of alternating materials).  This topic is of interest to a diverse array of physical settings ranging from ferroelectric perovskites \cite{dimer81,ferro} and polymers \cite{polymer} to optical waveguides \cite{sukh02}
and cantilever arrays \cite{sievers}.  Chains of beads provide a particularly interesting and versatile setting with which to study such problems because of their strong nonlinearity and the wide range of available material properties (and concomitant tunability) \cite{coste97,dar05,dar06,dar05b,nesterenko2}.  

Our aim in the present work is to investigate solitary wave propagation in 
chains of granular dimers \cite{anne05} combining experiments, numerical simulations, and 
theoretical analysis.  
Some preliminary results on dimer chains are discussed in \cite{nesterenko1}.  In this Rapid Communication, we consider a broad class of configurations consisting of stainless steel:PTFE (polytetrafluoroethylene) \cite{dar05,dar05b} dimer chains with different periodicities (obtained by varying the number of consecutive steel 
particles). We report very good agreement between experiments and 
numerics.  We also apply a long-wavelength approximation to the 
nonlinear lattice model to obtain a quasi-continuum 
nonlinear partial differential equation (PDE) description of the system.  Because of the chain lengths and the pulse amplitudes considered in our experimental and numerical analyses, the pulse propagation can be safely assumed to be within the highly nonlinear regime and the effects of gravity in the theoretical analysis have been neglected, in accordance with the discussion of \cite{hong01}. 
 We obtain analytical expressions for wave solutions of the PDE and find 
very good agreement between the widths and propagation speeds of these solutions with those 
obtained from experiments and numerical simulations.



\begin{figure}[tbp]
\centering \includegraphics[width=8.0cm]{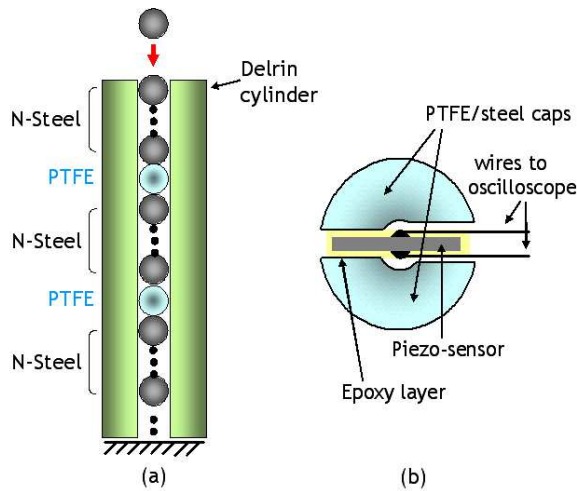}
\caption{(Color online) (a) Experimental setup for dimer chain consisting of a periodic array of $N$ consecutive stainless steel beads interspersed with 1 PTFE bead.  (b) Schematic diagram of the composition of the sensors placed in the chain.}
\label{setup}
\end{figure}

{\it Experimental Setup}.  The experimental dimer
chains were composed of vertically aligned beads in a delrin guide that contained slots for sensor connections.  
Each ``$N:1$ dimer" (see Fig.~\ref{setup}) 
included a variable number $N \in \{1,\dots, 7\}$ of the
high-modulus, large mass stainless steel beads (non-magnetic, 316 type) 
alternating with a single low-modulus, small mass PTFE bead in a 
periodic sequence.  The diameter of all spheres was 4.76 mm, and the number of beads in the chain was $38$.
Three piezo-sensors were embedded inside the particles in the system, 
as described in \cite{dar05,dar05b,dar06}.  The calibrated 
sensors (RC $\approx 10^3$ $\mu$s) were connected to a four-channel 
Tektronix Oscilloscope (TKTDS 2014), allowing direct visualization of the 
propagating pulse via force versus time curves and time-of-flight 
calculations of the pulse speed. Waves were generated by a striker 
(a $0.45$ g stainless steel bead) dropped from various heights.
The steel beads had mass 0.45 g,
elastic modulus 193 GPa, and Poisson ratio 0.3 \cite{metals,316}; 
and the PTFE beads had mass 0.123 g,
elastic modulus 1.46 GPa, and Poisson ratio 0.46 \cite{dar05,dupont}.  Previous studies of dimer chains only considered materials with similar elastic moduli \cite{nesterenko1}.



{\it Numerical simulations}.  We model a chain of $n$ spherical beads as a 
1D lattice with Hertzian interactions between beads \cite{nesterenko1}:
\begin{align}
	\ddot{y}_j &= \frac{A_{j-1,j}}{m_j}\delta_{j}^{3/2} - \frac{A_{j,j+1}}{m_j}\delta_{j+1}^{3/2} + g \,, \notag \\
	A_{j,j+1} &= \frac{4E_{j}E_{j+1}\left(\frac{R_jR_{j+1}}{R_j + R_{j+1}}\right)^{1/2}}{3\left[E_{j+1}\left(1-\nu_{j}^2\right) + E_j\left(1-\nu_{j+1}^2\right)\right]}\,,	
\end{align}
where $y_j$ is the coordinate of the center of the $j$th particle, $j \in \{1,\cdots,n\}$, $\delta_j \equiv \mbox{max}\{y_{j-1} - y_{j},0\}$ for $j \in \{2,\dots,n\}$, $\delta_1 \equiv 0$, $\delta_{n+1} \equiv \mbox{max}\{y_{n},0\}$, $g$ is the gravitational acceleration, $E_j$ is the Young's modulus of the $j$th bead, $\nu_j$ is its Poisson ratio,
$m_j$ is its mass, and $R_j$ is its radius.  The particle $j = 1$ 
represents the steel striker, and the $(n+1)$st particle represents the wall (i.e., an infinite-radius particle that cannot be displaced).  The initial
velocity of the striker is determined by experiments and all other particles start at rest in their equilibrium positions.

\begin{figure}[tbp]
\centering 
\centering \includegraphics[width=8.0cm]{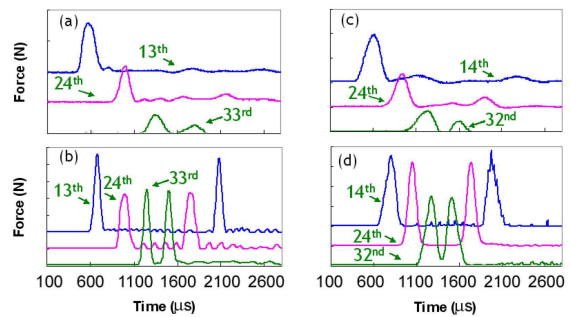}
\caption{(Color online)  Force versus time response obtained from chains of dimers consisting of (a,b) 1 or (c,d) 2 stainless steel beads alternating with 1 PTFE bead.  Panels (a,c) show experimental results, and (b,d) show the corresponding numerical data.  
The striker has a velocity of 1.37 m/s on impact and the $y$-axis scale is 2 N per division.  The numbered arrows point to the corresponding particles in the chain.  In both configurations, the second curve (showing the results for particle 24) represents a PTFE bead and the other curves represent steel beads.
}
\label{oneone}
\end{figure}


In Fig.~\ref{oneone}, we show the very good agreement 
between experimental and numerical results 
for $1:1$ [panels (a,b)] and $2:1$ [panels (c,d)] dimers of steel:PTFE particles. 
For the $1:1$ dimers, the dynamics indicate that the initial excited impulse develops into a solitary wave within the first 10 particles of the chain.  We also obtain robust pulses for $N:1$ dimers with $N >1$ (with $N$ as high as 7 in experiments and as high as 22 in numerics), though the transient dynamics and spatial widths of the developed solitary-like waves are different.

{\it Theoretical Analysis}. We focus our analysis on the prototypical $1:1$ dimer chain with beads of different masses (denoted $m_1$ and $m_2$). The 
rescaled equations of motion (without gravity; see the discussion below) can be written \cite{nesterenko1}
\begin{align}
	m_1 \ddot{u}_j &= (w_j-u_j)^k-(u_j-w_{j-1})^k\,, \label{eq1} \\
	m_2 \ddot{w}_j &= (u_{j+1}-w_j)^k - (w_j-u_j)^k\,, \label{eq2}
\end{align}
where $u_j$ ($w_j$) denotes the displacement of the 
$j$th sphere of mass $m_1$ ($m_2$). We consider a general power-law 
interaction to illustrate the comprehensiveness of our approach and use Hertzian 
contact ($k = 3/2$) to compare our theoretical analysis with our 
numerical and experimental results.  

The distance from $u_j$ to $w_j$ (from $u_j$ to $u_{j+1}$) is $D$ 
($2 D$).  Using $D$ as a small parameter, we develop a long-wavelength approximation (LWA) by 
Taylor-expanding Eqs.~(\ref{eq1})-(\ref{eq2}), for which we express $u_{j+1}$ 
($w_{j-1}$) as a function of $u_j$ ($w_j$). The resulting PDEs
need to be ``homogenized'' between the two ``species''. To accomplish this, we follow
\cite{pnevmatikos}, postulating a ``consistency condition'' between the two fields:
\begin{align}
	w & =\lambda \left(u + b_1 D u_x + b_2 D^2 u_{xx} + b_3 D^3 u_{xxx} \right. \notag \\
		&\quad \left.+ b_4 D^4 u_{4x}
+ \dots \right)\,. \label{eq3}
\end{align}
We then self-consistently determine the coefficients $\lambda$ and $b_i$ by demanding that Eqs.~(\ref{eq1}) and (\ref{eq2}) are identical {\it at each order}.
(Note that the subscripts in the LWA denote derivatives.)  The parameter 
$\lambda$ can take the values $1$ (for acoustic excitations) or $-m_1/m_2$ (for optical ones).  The nature of our experimental initial conditions {\it generically} 
leads to in-phase waveforms, so we restrict our 
considerations to $\lambda=1$ hereafter \footnote{To investigate optical excitations, it would be necessary to employ a fundamentally different and more complex framework involving a single precompressed bead and excitations of prescribed frequency in the corresponding linear-spectrum
gap of such a setting.}.  Considering the ensuing PDE at orders $D^k$--$D^{k+3}$, we 
find that $b_1=1$, $b_2=m_1/(m_1+m_2)$, $b_3=(2 m_1-m_2)/[3 (m_1+m_2)]$,
and $b_4=m_1 (m_1^2-m_1 m_2 + m_2^2)/[3 (m_1+m_2)^3]$.  Observe that 
these results correctly capture the uniform-chain limit of $m_1=m_2$, for which 
$w$ should represent the Taylor expansion of radius $D$ around $u$.

The resulting PDE,
\begin{align}
	u_{\tau\tau} &= u_x^{k-1} u_{xx}+ G u_x^{k-3} u_{xx}^3 + H u_x^{k-2} u_{xx} 
u_{xxx} \notag \\
	&\quad + I u_x^{k-1} u_{4x}\,,
\label{eq4}
\end{align}
has the same form (with different coefficients) as that obtained for uniform chains \cite{nesterenko1}.  In Eq.~(\ref{eq4}), $\tau=t \sqrt{2 k D^{k+1}/(m_1+m_2)}$ is a rescaled time, 
$G=D^2 (2-3 k + k^2)m_1^2/[6 (m_1+m_2)^2]$, 
$H=2 D^2 (k-1) (2 m_1^2+m_1 m_2-m_2^2)/[6 (m_1+m_2)^2]$, 
and $I=2 D^2 (m_1^2-m_1 m_2+m_2^2)/[6 (m_1+m_2)^2]$. We seek 
traveling-wave solutions $u\equiv u(\xi)$, with $\xi=x-V_s t$ (and $V_s = \frac{dx}{d\tau}$), and 
obtain an ordinary differential equation (ODE) for $u_{\xi}=v$.  We then 
change variables with the transformation $v=z^p$, where the power $p$ is 
chosen so that terms proportional to $z^{p-3} z_{\xi}^3$ disappear in 
the resulting ODE for $z = z(\xi)$.  Finally, an integrating factor 
$z^a$, with $a=1-k p + 3 (p-1) + p H/I$, converts the ODE to a tractable form,
$z_{\xi\xi}=\mu z^{\eta} - \sigma z$ 
where $\mu=V_s^2/(I (p+a))$, $\eta=1+p-k p$, and $\sigma=1/[I (kp + a)]$.  
Direct integration then yields:
\begin{align}
	u_{\xi} \equiv v \equiv z^p= B \cos^{\frac{2}{k-1}}(\beta \xi) \,,
\label{eq6}
\end{align}
with $B=\left(\mu/[\beta^2 s (s-1)]\right)^{1/(k-1)}$, 
$\beta=\sqrt{\sigma} (1-\eta)/2$, and $s=2/(1-\eta)$.  

The existence of such a trigonometric solution in the nonlinearly 
dispersive LWA suggests the possibility of finite-width
solutions that essentially consist of a single arch of the 
profile of Eq.~(\ref{eq6}), similar to what has been derived for uniform chains \cite{nesterenko1}.  Among the most notable properties of such solutions that are testable both experimentally and numerically are the amplitude-velocity scaling $B \sim V_s^{2/{(k-1)}}$ (which is similar to the single-species 
case \cite{nesterenko1,nesterenko2}) and the solution width 
$\pi/\beta$ (which explicitly depends on the mass ratio).  We have 
obtained an analytical expression for the width that is valid for all $k$, but the formula is too 
lengthy to write explicitly in the general case.  The expression for 
$\beta$ in the case of Hertzian interactions (i.e., $k=3/2$) is 
$\beta=(30-8 \omega+8 \omega^2)^{-1} (\sqrt{3} (8-5 \omega+5 \omega^2-\sqrt{4-4\omega+13\omega^2-18 \omega^3+9 \omega^4})
(((1+\omega)^2(15-4\omega+4\omega^2))/(D^2 (34-42 \omega+59 \omega^2-34
\omega^3+17 \omega^4+(-8+5 \omega-5 \omega^2)\sqrt{4-4\omega+13\omega^2-
18 \omega^3+9 \omega^4})))^{1/2})$, where $\omega=m_2/m_1$.  
This result, which is one of the main findings of the present work, 
generalizes all previously known limiting cases -- namely $m_1=m_2$, 
for which $\beta=\sqrt{10}/(5 D)$ [resulting in pulses extending to 
$5\pi /\sqrt{10} \approx 5$ sites], and $m_1 \gg m_2$ [resulting in 
pulses of $\sqrt{10} \pi \approx 10$ sites (composed of 5 cells 
with 2 sites each)] \cite{nesterenko1}.


\begin{figure}[tbp]
\centering
\includegraphics[width=8.0cm]{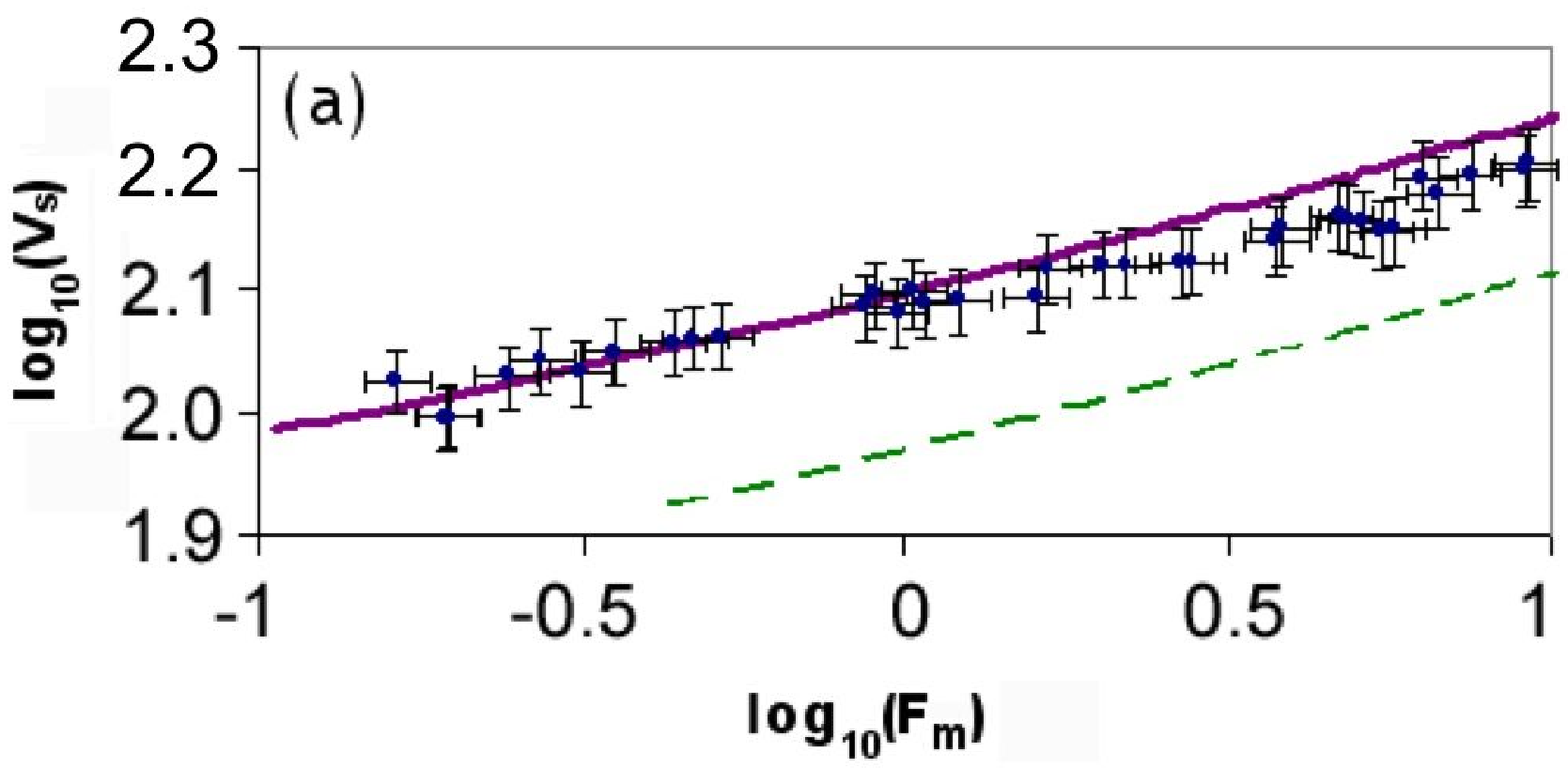}
\includegraphics[width=8.0cm]{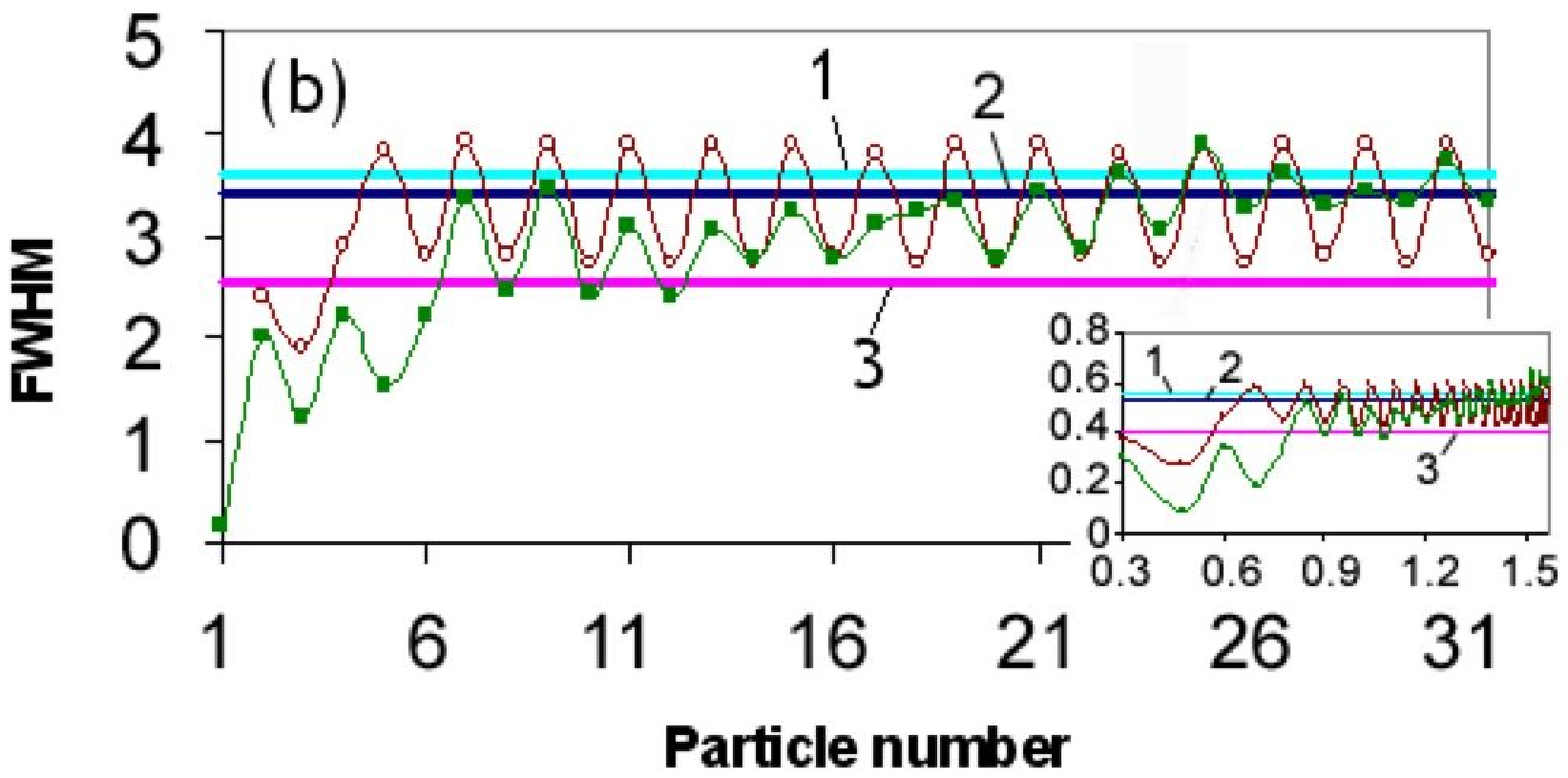}
\caption{(Color online) Comparison of experiments, numerical simulations, and theory.  (a) 
Scaling of the maximum dynamic force $F_{m}$ versus propagation velocity $V_s$ in a 1:1 stainless steel:PTFE chain.  The numerical simulations are shown by the solid curve and the experimental results are shown by points (with error bars).  The dashed curve
shows the numerical results using the elastic modulus $E = 0.6$ GPa for PTFE, the nominal static value reported in most of the literature \cite{dar05,dupont}. (b) Evolution of solitary wave width (full width at half maximum, or FWHM) as a function of bead number (also shown as a $\log$-$\log$ plot in the inset).  The experimental values are shown by solid (green) squares 
and the numerical values are shown by open (red) circles.  (In both cases, we include 
curves between the points as visual guides.)  The theoretical value for the 
FWHM with $m_1 \gg m_2$ is given by line (1), that for the 1:1 steel:PTFE chain is given by (2), and that for a homogeneous chain is given by (3).  
}
\label{compare}
\end{figure}


{\it Comparison between experiments, numerical simulations, and theory}.  To begin, we examine the 
scaling of the maximum dynamic force $F_{m}$ of the solitary waves versus their propagation speed $V_s$.  The theory predicts that $F_{m} \sim B^k \sim V_s^{2k/(k-1)}$, which for $k = 3/2$ yields $V_s \sim F_{m}^{1/6}$.  We tested this experimentally using a 1:1 steel:PTFE dimer chain 
by dropping the striker (a stainless steel bead) from initial heights ranging from 0 m to 
1.2 m. We placed sensors in beads 15 and 34 and measured the corresponding peak force.  We averaged their amplitudes [$F_{m} = (F_{15}+F_{34})/2$] and  
obtained the corresponding wave speed ($V_s$) using time-of-flight 
measurements.  We calculated these diagnostics similarly in our numerical simulations. 
As shown in Fig.~\ref{compare}(a), we obtain very good agreement between numerics and experiments.  A least-squares fit of the numerical simulations (using the experimental configuration of 38 beads and removing gravity) yields
$V_s \sim F_{m}^{0.1666}$, in excellent 
agreement with the theory (despite the small number of particles).  

To better highlight some of the interesting dynamics of dimer chains, we examined the evolution of the solitary-wave width (the full width at half maximum, or FWHM) as the wave progresses down the chain of beads.  As shown in Fig.~\ref{compare}(b), the numerical and experimental results are in 
very good agreement with each other and with the ``homogenized'' theory (which
should be expected to capture the average of the relevant oscillations).  The $F_m$ values in the numerics and experiments alternate from one bead to another because consecutive beads are composed of different materials.
As one can see from Fig.~\ref{compare}(b), the average FWHM of the dimer is decidely different 
from that of a homogeneous chain ($m_1 = m_2$) and even from that of 
the $m_1 \gg m_2$ limiting case. 
It is best captured by the theoretical calculation above (for the proper
ratio of masses), clearly evincing the relevance of our theoretical 
approach. As a quantitative measure, the relative error of the theoretical
prediction versus the computational-average FWHM is $22.8\%$ for $m_1=m_2$, 
$9.2 \%$ for $m_1 \gg m_2$, and $2.5 \%$ for the relevant mass ratio.  In conjunction with the theoretical analysis developed in this paper, our experiments and numerics reveal that the equilibrium conditions (in a horizontal chain) for the wave width are achieved even at very short propagation distances (i.e., with a very small number of beads).  For vertical chains of beads that contain a large number of particles (several hundred or more) or are excited by smaller-amplitude pulses, the presence of the nonuniform gravitational precompression should be taken into account.  As discussed in \cite{hong01}, this leads to a 1/3 power law scaling in the wave width as a function of particle number.

%
%


 {\it Conclusions}.  We examined the propagation of solitary waves in heterogeneous, periodic chains of granular media using experiments, numerical simulations, and theory.  Using different periodicities, we found that such heterogeneous systems robustly support the formation and propagation of highly localized nonlinear solitary waves (with appropriate widths that depend on the periodicity).  We used force-velocity scaling (which is the same as for homogeneous chains) and solitary-wave width (which depends on the mass ratio of the dimer materials) as relevant benchmarks for the very good agreement between our three approaches. This qualitative and quantitative understanding of the dimer dynamics also paves the way for studies in increasingly heterogeneous media (e.g., trimers) in both one and higher dimensions, as well as of more complex (optical) modes in such systems.


 
{\it Acknowledgements} We acknowledge support from the Caltech Information Science and Technology initiative (M.A.P.), start-up funds from Caltech (C.D.), and NSF-DMS and CAREER (P.G.K.).  I.S. performed his work as a visiting student of GALCIT at Caltech.  We 
thank A. Molinari and V.~F. Nesterenko for useful discussions.

 
 \vspace{.05 in}
 
$^*$ Electronic address: {\it daraio@caltech.edu}

 \end{document}